# Preliminary Investigation of the Frictional Response of Reptilian Shed Skin

H. A. Abdel-Aal[1]*, R. Vargiolu[2], H. Zahouani[2], M. El Mansori[1]

1. *Arts et Métier ParisTech,LMPF-EA4106, Rue Saint Dominique BP 508, 51006 Chalons-en-Champagne, France*
2. *Laboratoire de Tribology et Dynamique des Systèmes, UMR CNRS 5513, ENI Saint – Etienne - Ecole Centrale de Lyon .36 Avenue Guy de Collongue, 69131 Ecully cedex. France*
*corresponding author* hisham.abdel-aal@ensam.eu

## Abstract

Developing deterministic surfaces relies on controlling the structure of the rubbing interface so that not only the surface is of optimized topography, but also is able to self-adjust its tribological behaviour according to the evolution of sliding conditions. In seeking inspirations for such designs, many engineers are turning toward the biological world to correlate surface structure to functional behavior of bio-analogues. From a tribological point of view, squamate reptiles offer diverse examples where surface texturing, submicron and nano-scale features, achieve frictional regulation. In this paper, we study the frictional response of shed skin obtained from a snake (Python regius). The study employed a specially designed tribo-acoustic probe capable of measuring the coefficient of friction and detecting the acoustical behavior of the skin in vivo. The results confirm the anisotropy of the frictional response of snakes. The coefficient of friction depends on the direction of sliding: the value in forward motion is lower than that in the backward direction. In addition it is shown that the anisotropy of the frictional response may stem from profile asymmetry of the individual fibril structures present within the ventral scales of the reptile.

## Nomenclature

| | |
|---|---|
| $A_{VS}$ | Area of a ventral scale (defined in figure 3-b) mm$^2$ |
| H | Head |
| Pa | Radiated acoustic power (W) |
| Ra | Mean arithmetic value of roughness (µm) |
| Rq | Root mean square average of the roughness profile ordinates (µm) |
| Rz | Mean peak-to-valley height of the roughness (µm) |
| T | Tail |
| dB | Decibels |
| $l$ | Length of an individual fibril (µm) |
| **Directions** | |
| AE-PE | Anterior Posterior |
| DB | Diagonal backward |

| DF | Diagonal forward |
| LR | Lateral right hand side |
| LL | Lateral left hand side |
| LF | Lateral forward, |
| LB | Lateral backward. |
| SB | Straight backward |
| SF | Straight forward |
| *Acronyms* | |
| DFE | Differential Friction Effect |
| COF | Coefficients of friction |
| VSAR | Ventral Scale Aspect Ratio |
| FAR | Fibril Aspect Ratio |
| RSP | Reference Sound Pressure ($\mu$m) |
| SPL | Sound pressure level |

*Greek symbols*

$\lambda$     Separation distance between rows of fibrils

$\rho$     Fibril density (fibril/mm$^2$)

## Introduction

Friction is a major concern in the design of moving parts. It is a principal source of energy losses. Considerable amounts of the input energy to a moving assembly may be lost to combat friction. These losses take several forms (heat, acoustical emission, etc,). In addition, in most cases rubbing materials endure mass loss as a consequence of sustained rubbing (i.e., wear). Wear has many detrimental influences on the structural integrity of any rubbing pair. One question that is frequently encountered in the design of rubbing pairs is how to limit the frictional losses and to minimize, if not prevent, wear from occurring. There have been numerous proposals that address such a goal. Recent paradigms, however, are shifting toward custom engineering of surfaces as a fundamental step in the struggle to control friction and wear. Customization of both texture and topography are intended to tackle the problem at the primitive level of asperity contact. Therefore, at least in theory, customized surface topography should facilitate control of the rubbing process which, potentially, may keep energy losses and material damage within an acceptable threshold.

Many researchers envision so called "deterministic surfaces" as the future of surface engineering [1, 2]. These surfaces not only would be of optimized topography but also their function would be an integral feature of the overall function of the system they enclose. Engineering these surfaces for rubbing applications depends on understanding the interaction between surface topography, roughness parameters and frictional response. A structured surface for enhanced tribological performance, ideally, should be capable of tuning its frictional response, when needed, to optimize system function. In nature, there is an abundance of hierarchically-structured naturally occurring surfaces that deliver similar functionality [3]. The diversity and richness of these examples have fuelled interest in the study of natural surface designs for mimicry, or for deduction of surface design rules applicable to deterministic constructs.

Two topics are of particular interest when studying naturally occurring surfaces for tribological purposes. The first is the mechanics of interaction between surface topography and friction response. The second is the role of surface topological features in tuning tribological behavior in response to changes in the contacting terrain. A rich resource for such a purpose is the squamata order of reptiles.

Squamata, or the scaled reptiles, comprise the largest order of reptiles. They are a major component of the world's terrestrial vertebrate diversity [4]. Squamata has been traditionally classified into three suborders: Lacertilia (lizards) with 16–19 families, Serpentes (snakes, around 3000 species) with 14–17 families and Amphisbaenia (worm lizards) with 3–4 families [5]. They have diversified on all major continents, and occupy a remarkable diversity of ecological niches [6]. For frictional studies, the so-called legless (or limbless) locomotion, which is particular to snakes, is of special interest. This is because the complex role that the skin of a snake assumes in the different modes of motion that the reptile exhibits is similar to that of a tribo-surface.

Snakes use several modes of terrestrial locomotion (simple undulation, lateral undulation, side winding, concertina locomotion, slide-pushing, and rectilinear locomotion): A particular mode of locomotion a snake adopts in a given instance is a function of several factors (e.g., the contact interface, speed, temperature, etc.,). Although there are distinct kinematic differences between the individual modes of motion, they all share their origin in muscle activity. Transfer of motion between the active muscle groups and the contacting substrate will thus depend on generation of sufficient tractions. Generation of tractions and accommodation of motion is handled through the skin. The skin of a snake while transferring locomotion tractions also has to accommodate the energy consumed in resisting the motion.

Muscular activity for locomotion comprises sequential waves of contraction and relaxation of appropriate muscle groups. The number, type, and sequence of muscular groups responsible for the initiation, and sustainment, of motion, and thus employed in propulsion vary according to the particular mode of motion. This implies that contact stiffness in dynamic as well as in static friction constantly varies. Generation of tractions for motion also depends on the habitat and the surrounding environment. This, in turn, affects the effort invested in initiation of motion and thereby affects the function of the different parts of the skin. Different regions of the skin will have different functional requirements. Moreover, the life habits of the particular species (e.g., defense, preying, swallowing, camouflage, etc.,) will require different topographical features within the different skin regions. These aspects render snakes ideal study objects when the role of surface features in controlling friction is considered.

Previous work by some of the current authors has established the multi-scale nature of the topographical features of snakeskin [7]. The authors suggested that selective engagement of the appropriate scale of the surface topographical features allows for regulation of friction. However, no friction measurements were undertaken. In this work, therefore, we report initial results of a study designed to characterize the dynamical frictional behaviour of snakes. The species chosen for this work is the python regius (royal python or ball python). Shed skin obtained from the snake was subject to frictional measurements in several directions that mimic the principal modes of locomotion of the reptile. The focus was on the ventral side of the skin as it is the principal sliding part in snakes.

## 2.   Species description

Python Regius.is a constrictor non-venomous snake. It typically inhabits Africa. The species has the alternative name "ball-python" due to the reaction of the reptile upon fear or distress. In such a case, it assumes a spherical position with the head tucked under the trunk, for protection.

Figure (1) depicts the general features of the skin. Dorsal skin comprises several dark and light colored blotches. The ventral skin, on the other hand, is mainly cream in color with occasional black markings. Details of the micro features of the scales on the dorsal and ventral sides are depicted in the figure as well. Two SEM pictures, at different magnifications (x=1000 and x=10000), are provided for each skin color. As shown, scale micro-features comprise fibrils arranged in rows. The rows are arranged in waves rather than in straight lines. The shape of the fibrils, and the spacing between fibril waves, appear to be inconsistent. They vary by location and by color of skin. Fibril tips point toward the posterior end of the reptile. The shape of fibril tips also displays some variation with color of skin. In the dark-colored dorsal scales, fibrils are tapered and have a sharp tip. Scales within the bright colored and the ventral regions exhibit fibrils of a more rounded tip and of uniform width throughout the fibril length. Moreover, the density of the fibrils seems to be different within the different color regions (denser within the dark colored region).

The build of the snake is non-uniform. The head-neck region as well as that of the tail, is thinner than the-region containing the trunk. The trunk meanwhile is the region of the body where most of the snake body mass is concentrated. It is more compact and thicker than other parts. Consequently, most of the load bearing upon sliding occurs within the trunk region. The tail section also is rather conical in shape, although considerably thinner than the trunk. The overall cross section of the body is quasi elliptical rather than circular (the circumference of the upper half of the cross section is lengthier than the circumference of the lower half), as shown schematically in figure 2. A small segment of the body bears the load while sliding. This is shown in figure 2 as line v-v. As shown in figure, this segment longitudinally contains a band of ventral scales. The ventral side of the reptile therefore bears the reaction of the terrain and is responsible for the frictional response of the reptile and the generation of locomotion tractions. As such, discussion from this point onwards will focus on the geometrical and metrological features of the ventral scales.

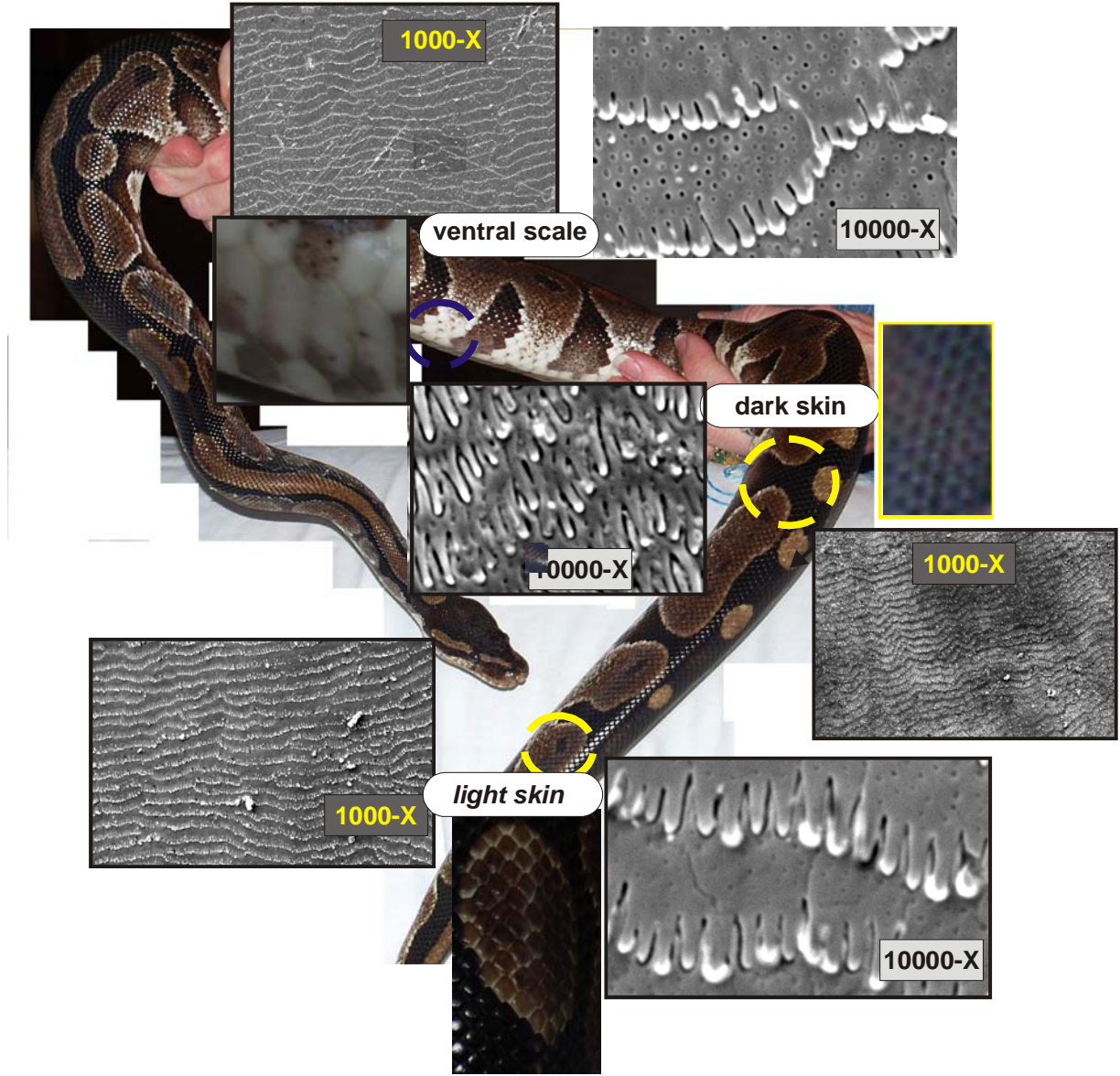

*Figure 1. General appearance of the Python Regius and SEM details of the three skin colors of the species: Dorsal skin light colored, Dorsal skin dark colored and the Ventral skin. All observations were performed on a JEOL JSM-5510LV SEM using an acceleration voltage that ranged between 4 kv ≤ V ≤ 6 Kv).*

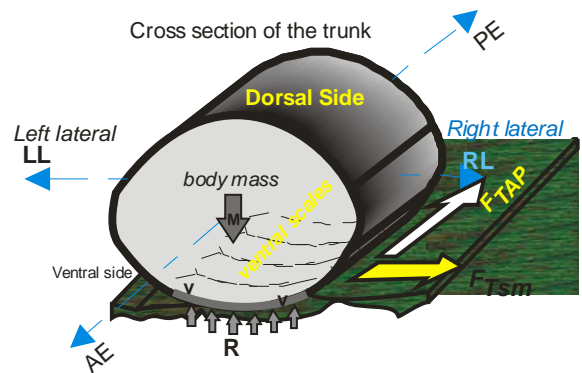

*Figure 2: schematic of the cross sectional shape of the trunk region (right hand side) and its interaction with the environment. Line v-v represents the segment of the trunk responsible for accommodation of frictional tractions in forward motion ($F_{TAP}$) and in side-motion ($F_{Tsm}$). Line v-v is the projection of the width of the band that contains the ventral-scales. Note the scale of the width compared to the circumference of the trunk of the reptile.*

## 3. Structure of Snake skin

Scales totally cover the body of a snake. Scales are essentially skin folds of various sizes and shapes (mainly polygons). The function of scales is multifaceted. They can perform biological functions such as helping to retain moisture and exchange heat with the surrounding environment. They also may aid in concealing the reptile for defense or for hunting purposes. The number of scales in any given snake is fixed. The reptile hatches with a constant number of scales that is specific to the particular species. Scale counts do not change with aging. Rather, the scales increase in size to accommodate growth. Scales also include the ornamentation, texturing, of the skin. Ornamentation is thought to alter the characteristics of the reptile surface. Snakes undergo a shedding process through which their entire skin is replaced. This process occurs periodically 4 to 8 times a year depending on the species [8].

Reptilian skin has two principal layers: the *dermis* and the *epidermis*. The dermis is the deeper layer of connective tissue. It contains a rich supply of blood vessels and nerves. The epidermis consists of up to seven sub-layers or "strata" of closely packed cells [9]. These form the outer protective coating of the body. The "epidermis" has no blood supply, but its' inner most living cells obtain their nourishment by the diffusion of substances to and from the capillaries at the surface of the "dermis" directly beneath them [10]. The epidermis has a layered composition. It has seven layers. The first, and deepest, of such layers is the *stratum germinativum*. The stratum germinativum is lined with cells that have the capacity for rapid division. Six additional layers form each "epidermal generation", (the old and the new skin layers). These are: the clear layer and the lacunar layer, which matures in the old skin layer as the new skin is growing beneath; the (α)–layer, the mesos layer and the (β)-layer, these layers consist of cells which become keratinized with the production of two types of keratin (α and β keratin). Finally, there is the "Öberchautchen" layer, which is the toughest outer most layer of keratinized dead skin cells. The "Öberchautchen", being the outermost layer of the body, is the layer that accommodates locomotion tractions and is the principal layer in any skin molt. The frictional behavior of this layer constituted the principal interest in the current work.

## 4  Materials and methods

### 4.1 Skin treatment

All observations reported herein pertain to shed skin obtained from five male Ball pythons (Python regius). The snakes were housed individually in glass containers. The received shed skin underwent a water assisted unfolding procedure. This procedure consisted of soaking the skin in distilled water at room temperature for two hours to unfold. Following soaking, the skin was dried by enclosing in paper towels for 4 hours. Finally, upon removal from the paper towels, the skin was cleaned by blowing compressed air stored in metallic canisters, wrapped in wax paper, and stored in sealed plastic bags until being used in the experiments.

### 4.2 Surface texture metrology

Measurement of surface texture was performed using a white light interferometer (WYKO 3300-3D automated optical profiler system). Samples used for profilometry didn't receive any treatment beyond the initial unfolding and cleaning procedure described earlier. All resulting White light Interferometry (WLI) were processed to extract the surface metrological parameters using two software packages: Vision $^{®}$v. 3.6 and Mountains$^{®}$ v 6.0.

For each chosen area, on a particular snake, a series of five SEM pictures at different locations within the particular scale were recorded. The pictures were further analyzed to obtain fibril geometric information (counts, distance between fibrils, and length of individual fibrils). The average (arithmetic mean) of the information sets of the selected ventral scales were then plotted against the non-dimensional distance.

### 4.3 Friction measurements

All measurements utilized a patented bio-tribometer. Complete description of the device is given elsewhere [11]. The device includes a tribo-acoustic probe that is sensitive to the range of friction forces and the acoustic emission generated during skin friction. It is also capable of measuring normal and tangential loads and of detecting sound emission due to sliding. The probe comprises a thin nitrocellulose spherical membrane, 40 mm in diameter, with a thickness of 1 mm. The probe material has a Young's modulus of 1 GPa. The roughness of the probe, $R_a$, is 4 μm and the mean value between peak to valley, $R_z$, is 31 μm.

In all frictional tests, the skin was stationary and the tribo-probe was moving at an average speed of 40 mm/s using a normal force of 0.4 (±0.05) N. The skin used in measurements consisted of 150 mm long patches taken from four locations on the ventral side of the shed skin. Skin samples didn't receive any chemical or physical treatment beyond the water-assisted unfolding procedure described in the previous section.

To mimic the effect of the body of the snake on the skin, before starting an experiment, the particular skin patch was placed on a rectangular elastic pad of dimensions length L= 200 mm ,width W=100 mm and thickness of approximately 4 mm. The pad is made of silicone rubber (Silflo®™, Flexico Developments Ltd., Potters Bar, UK). The mechanical characteristics of the material of the pads are: Young's Modulus $E = 2$ MPa at 20 °C, Poisson ratio $v = 0.3$, and the spring stiffness $K$ is 300 N/m. The same pad was used in all experiments.

## 5.    Results and discussion

### 5.1    Dimensional metrology of ventral scales

In this work, we consider the metrology of the ventral scales at two levels of measurements. The first is at the individual fibril level, which is in the order of few microns whereas the second is at the "entire scale" level, which is in the order of several millimeters.

Four parameters are used to describe the geometry of the fibrils.  These are the length of an individual fibril, $l$, fibril density per unit area of ventral scale, $\rho$,  the spacing between fibril rows $\lambda$, and the Fibril Aspect Ratio, FAR, defined as (see figure 3-a),

$$FAR = \frac{l}{w} \tag{1}$$

Where $w$ is the width of an individual fibril. These parameters, along with, the profile of the fibril tips, yield a complete description of the topography of the fibril microstructures within the ventral scales.

As mentioned elsewhere [12], fibrils of maximum length for the python regius were located within three regions: the top and the bottom boundaries of the trunk, and the mid-section. Their lengths were found to fall within the range $1.3 < l < 1.5$  µm.   The shortest fibrils meanwhile, were located within two regions: the head-neck region, and the trailing end of the load bearing volume. Their length was approximately 0.8 µm.

The distribution of fibril density ($\rho$) along the body was evaluated by counting the number of fibrils from SEM images taken at a magnification of X=10,000.  The results indicated that density of the fibrils has a non-uniform distribution that fell within the range $3*10^5$ fib/mm$^2 < \rho < 6*10^5$ fib/mm$^2$.  Additionally, fibril density was found to increase toward the middle section of the trunk and to be minimal within the non-load bearing zones of the trunk.

The distribution of the separation distance $\lambda$ along the body of the reptile was found to be non-uniform.  Internal spacing was larger within the trunk.  The maximum was located roughly around the mid-section.  The distance between fibril waves fell within the range 3.5µm $< \lambda < .5$ µm.  The shortest spacing (approximately 3.5µm) was roughly located within the non-load bearing portions of the body (i.e.; the head and tail sections) [13].  Using the dimensions of the fibrils it is possible to estimate the Fibril Aspect Ratio, FAR, from equation (1) as $1.5 < FAR < 2.4$.

As mentioned earlier, snakes hatch with a fixed number of scales.  For the Python regius species 208 scales are located within the ventral side [14].  For the snakes used in this work, the area of the ventral scales (defined in figure 3-b) fell in the range $75 < A_{VS} < 125$ mm$^2$ whereas, the Ventral Scale Aspect Ratio VSAR ranged between $2 < VSAR < 3.2$.

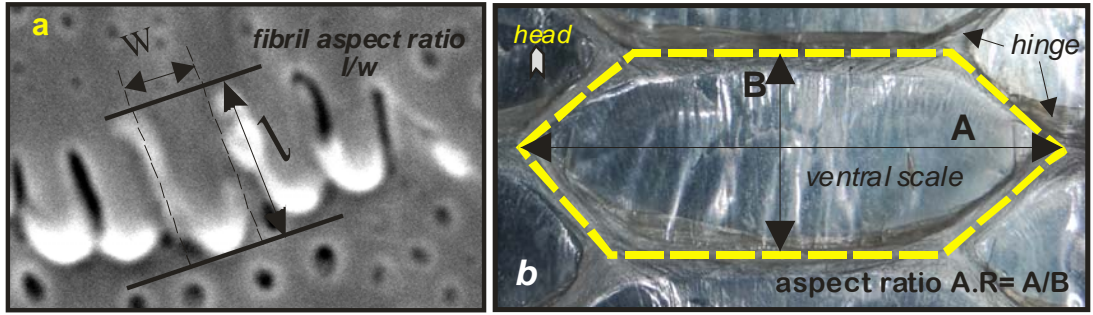

*Figure 3 Definition of the geometric parameters used for scale characterization, (a) definition of the fibril aspect ratio, and (b) definition of the surface area of the ventral scale, $A_{vs}$, and Ventral Scale Aspect Ratio (VSAR).*

## 5.2   Surface metrology of fibrils

The shape of fibril tips was considered important by Hazel *et.al*, [15], in invoking anisotropy of the frictional response of snakeskin. As such, special attention was given to the extraction of the shape of the tips. Profiles were extracted using two methods: processing of WLI and AFM scans of 40 µm by 40 µm regions on the ventral scales. AFM samples were scanned using a Park systems XE-150 cross functional AFM in topography mode. Software supplied by the manufacturer was used to extract metrological parameters from the scans.

Figure 4 (a-f) shows the extracted profiles. For WLI four directions for profile analysis were chosen. These are shown schematically in figure 4-a. As shown in figure the analysis proceeded along the two principal axis (the Anterior-Posterior axis AE-PE, and the lateral axis LR-LL) and two diagonal directions. The orientation of the figure mimics the actual contact situation of the fibrils. Thus, the bottom of the figure represents the sliding substrate, which the skin should establish contact with in actual sliding. The directions of motion forward and backward are designated according to whether motion proceeds from the anterior end (head) toward the posterior end (tail) of the reptile or the converse. The former is considered forward whereas the later is considered backward respectively. Based on the classification the edges of a fibril are designated as "*leading*" or "*trailing*". It is seen from the figure that for all directions, except in figure 4-d, which illustrates profile in the lateral axis, that the fibril tips are of asymmetric profile. In particular, the slope of the trailing edge is steeper than that of the leading edge. In the lateral direction, however, there is no clear distinction of the slopes. It seems that the fibril tips are rather symmetric in such a direction.

Asymmetry of tip profiles is apparent in figure 4-f. The figure presents a representative AFM scan of a square area (40 µm by 40 µm) of one of the ventral scales; the extracted profile is plotted under the scan image. The points designated as *m* and *n* denote the location of extracted profile data within the scanned area. As seen, the scan is oriented along the anterior-posterior axis. Again, asymmetry of fibril tip profile is noted. Note also that the distance m-n (spacing between two fibril rows) in the figure is in the order of 3 µm < m-n< 4 µm which is consistent with the range of row inter-spacing, λ, extracted from WLI.

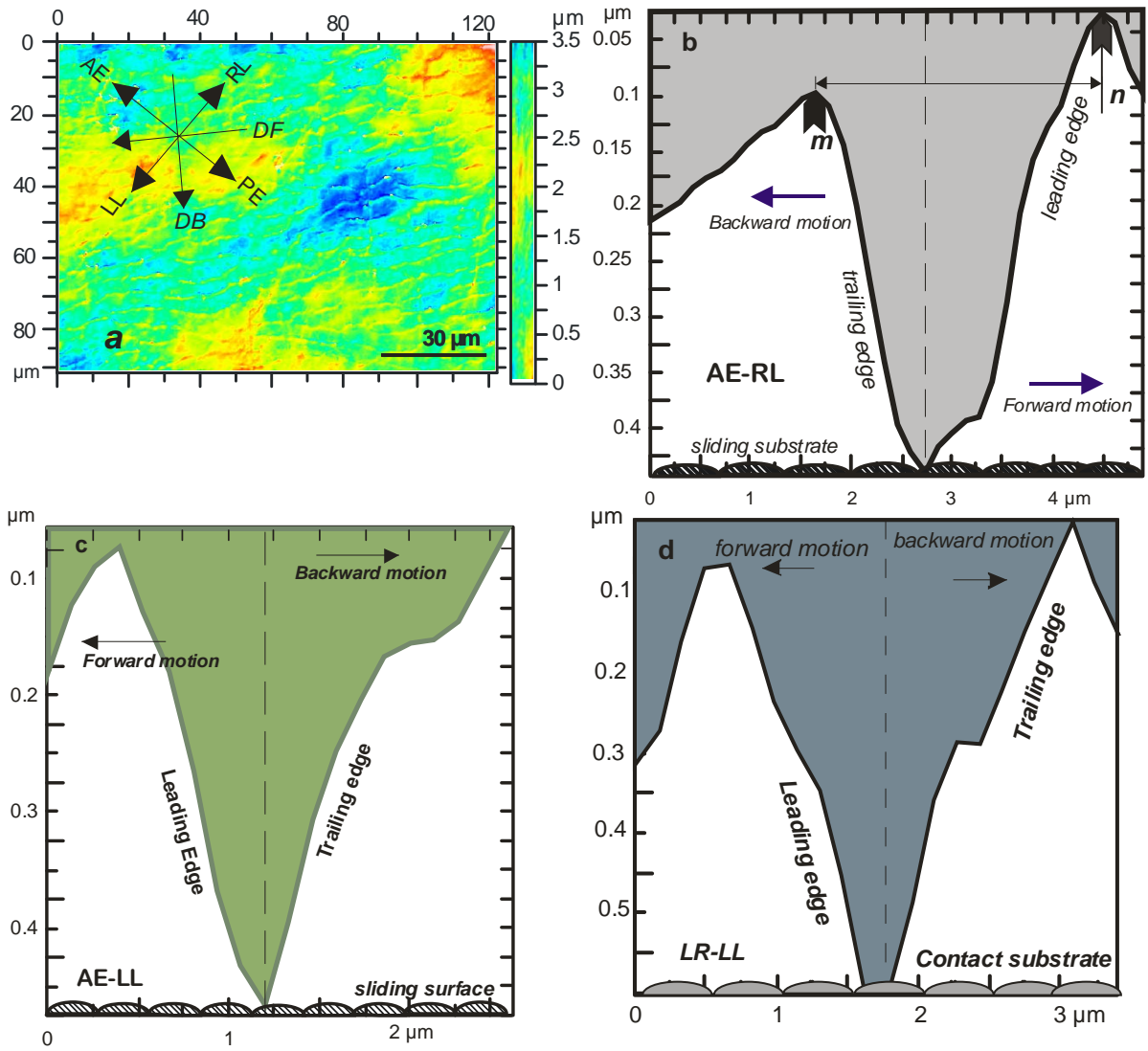

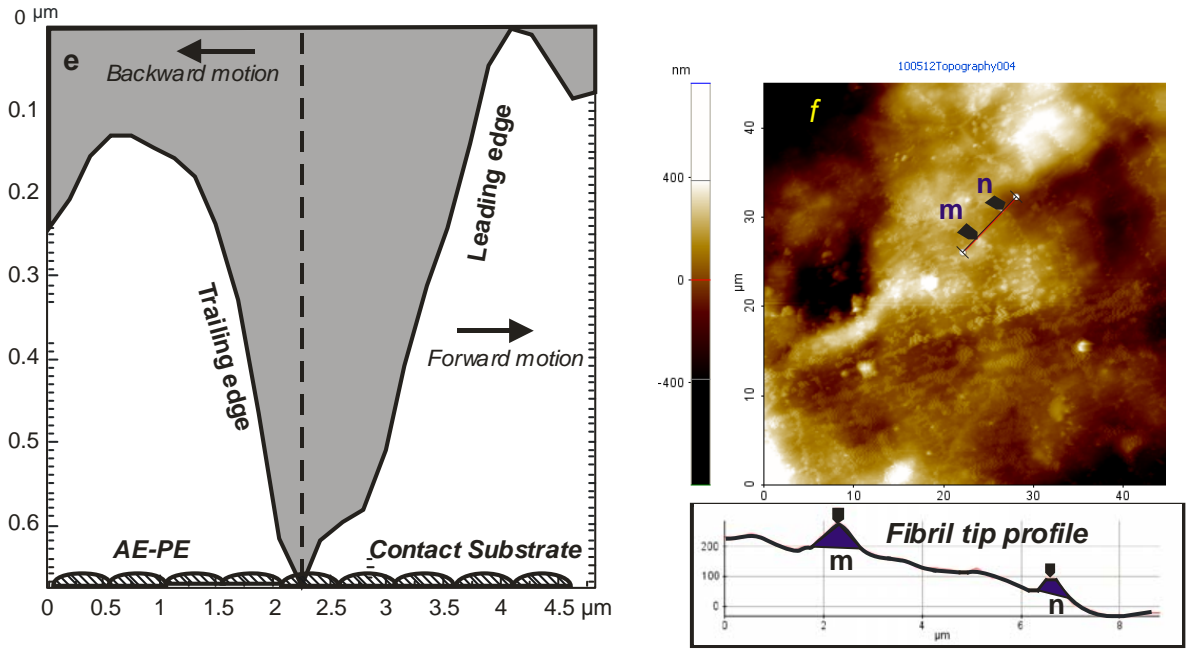

**Figure 4** *The shape profile of individual fibril tips in the ventral scales of the reptile: (a) illustration of the direction of analysis of WLI sown on a typical scan, AE Anterior end, PE Posterior end, RL lateral right, and LL is lateral left.  Note the orientation of the fibril waves with respect to the major axis of the reptile  (b) Profile of the fibril tip in the diagonal direction between the Anterior end and the Right Lateral axis, (c) Profile of the fibril tip in the diagonal direction between the Anterior end and the Left Lateral axis, (d) Profile of the fibril tip in the direction of the Lateral axis, (e) Profile of the fibril tip in the direction of the Anterior Posterior axis, (f) Shape of the tips obtained from AFM scanning, **m** and **n** are locations for further analysis by AFM.*

## 5.3    Sliding behavior
### 5.3.1   frictional behavior

For each skin patch used in the experiments, frictional measurements were performed in four directions.  The chosen directions, shown in figure 5 coincide, in general, with the major axis of the reptile.  Thus, measurements from the anterior end toward the posterior end of the reptile and the converse represent rectilinear motion.  This is denoted as straight forward (SF), and straight backward (SB) respectively in the figure. Measurements in side motion along the Lateral axis LR-LL is lateral forward (LF) and the converse is lateral backwards (LB). Measurements in diagonal motion forward (DF), and Diagonal backward (DB) were also performed.

Ten measurements were obtained in each direction. Figure (6-a) depicts the scatter of measurements taken in rectilinear motion along the AE-PE axis in the backward direction. The square symbol denotes the arithmetic mean of the ten readings. The repetition of the measured forces, in the normal and tangential directions, for the same sliding direction is shown on the right hand side of the page (figure 6-b).

Figure 7 presents a summary plot of the measured coefficients of friction, COF, in all the predetermined directions. Each value is an average of ten consecutive measurements taken at the same direction of sliding under identical loading conditions.

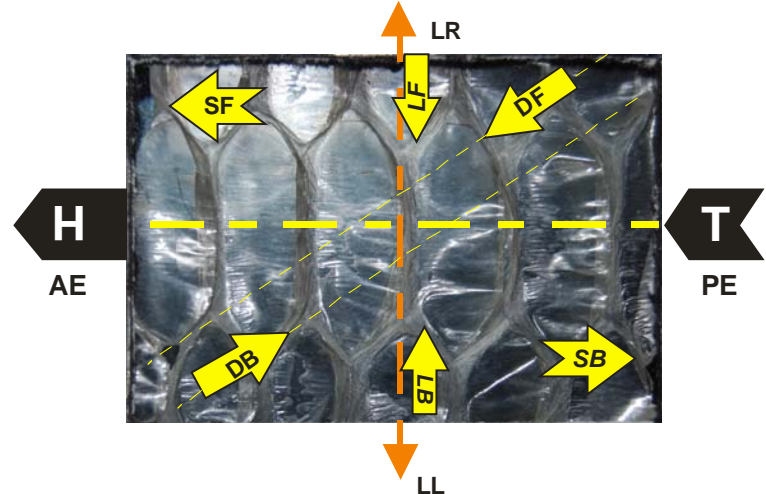

*Figure 5 Orientation of the frictional measurements performed on the skin with respect to the head and tail of the reptile. Straight motion denotes rectilinear locomotion along the Anterior Posterior axis (AE-PE). Forward direction is going from tail to head and backward is the converse. Similar convention is adopted for diagonal motion. For lateral motion going from the right hand (LR) side to the left hand side (LL) is considered forward and the converse is considered backwards. H-head, T, tail, SF-straight forward, SB-straight backward, DF-diagonal forward, DB-diagonal backward, LF-Lateral forward, and LB-lateral backward.*

Two plots are shown in the figure. Figure 7-a, is a plot of the coefficient of friction measured in rectilinear motion (forward and backward directions) compared to the value of the coefficient measured in diagonal motion (again forward and backward). Figure 7-b, meanwhile presents a plot of the COF in rectilinear motion compared to that obtained in lateral motion (forward and backward according to the convention of figure (5)).

The results show that friction in forward motion is, in general, less than that encountered in backward motion (observe the difference between the values obtained for the SF direction and that for the SB direction). Similar trend is noted for friction in the diagonal directions. The COF in the diagonal backward direction is greater than that in the diagonal forward direction.

The COF when moving diagonally differs from that in straight motion. Resistance to motion, however, from tail to head (i.e., forward) is less than that from head to tail (i.e., backward) regardless of the direction of motion (forward or diagonal). These general trends are in line with measurements obtained by other researchers [16, 17] who confirmed the anisotropy in friction as well as the increase in resistance to backward motion (although on different species).

The value of the friction coefficient in lateral motion doesn't show significant variation between forward motion (going from left to right) and backward motion (going from right to left) on the scale. Note that, while measurements in rectilinear and diagonal motion include passing over several scales, the measurement of the lateral coefficient of friction entails measurement over the same ventral scale. While this trend is consistent with the findings of other researchers, we present the values here with caution. This is because of the nature of the probe used in the experiments which requires a certain distance for the signal to reach a steady state. This distance may be equal to the length of the lateral axis of the particular scale. As such, while cautiously accepting the results, plans are underway to examine the accuracy of this measurement by applying an alternative technique.

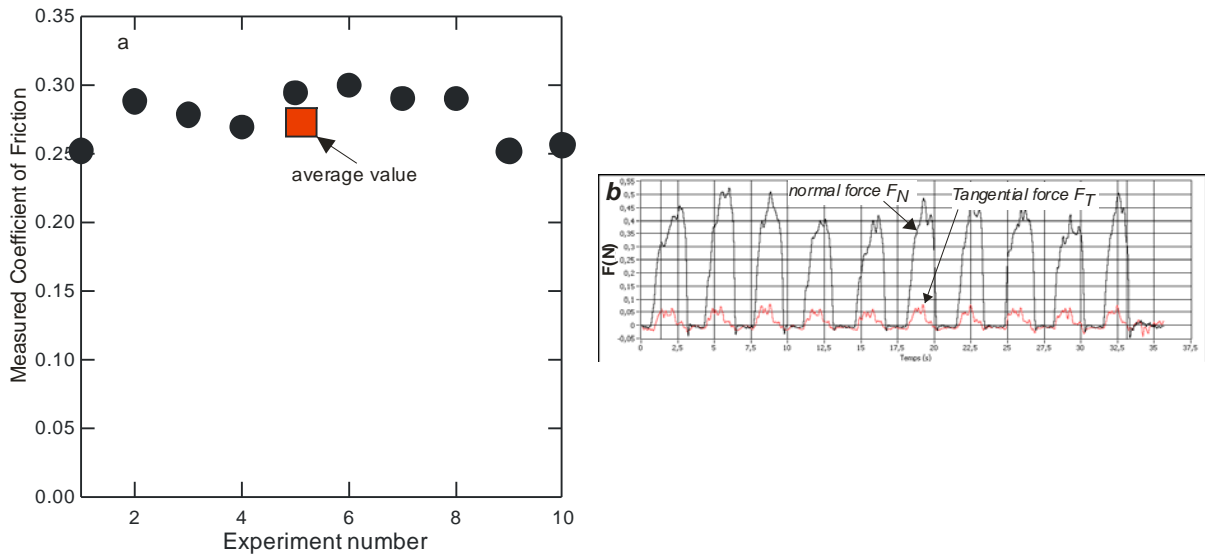

*Figure 6 Scatter of the data obtained for consecutive measurements in rectilinear backward motion (figure 6-a). Figure 6-b depicts plots of the normal and tangential forces in each of the measurements*

Detailed comparison of the results presented in this work and those of Berthe et al. [16] are not possible at this time for two reasons. Firstly, the results presented in this work pertain to dynamic measurements were as the result presented in the work of Berthe pertain to static measurements. Secondly, Berthe presented comprehensive data for localized measurement using a technique that fundamentally differs from the technique applied in this work so that a meaningful one to one comparison would not be objective. Nevertheless, the data presented in this work reflect a similar trend to that reported by Hazel [15] and Berthe [16]. We note however, that, in this work, *dynamic measurements* of the COF in the diagonal directions are reported for the first time in literature. The trend of the current results agrees with the findings of Zhang et, al [17] who examined the friction of ventral scales obtained from a Burmese python. Zhang's experimental result showed that the friction coefficient is different at different parts and different directions. He reported friction anisotropy of the skin more than 60% of forward motion with backward COF higher than forward one. In this work the friction coefficient in backward motion was found to be around 40% higher than that of forward motion.

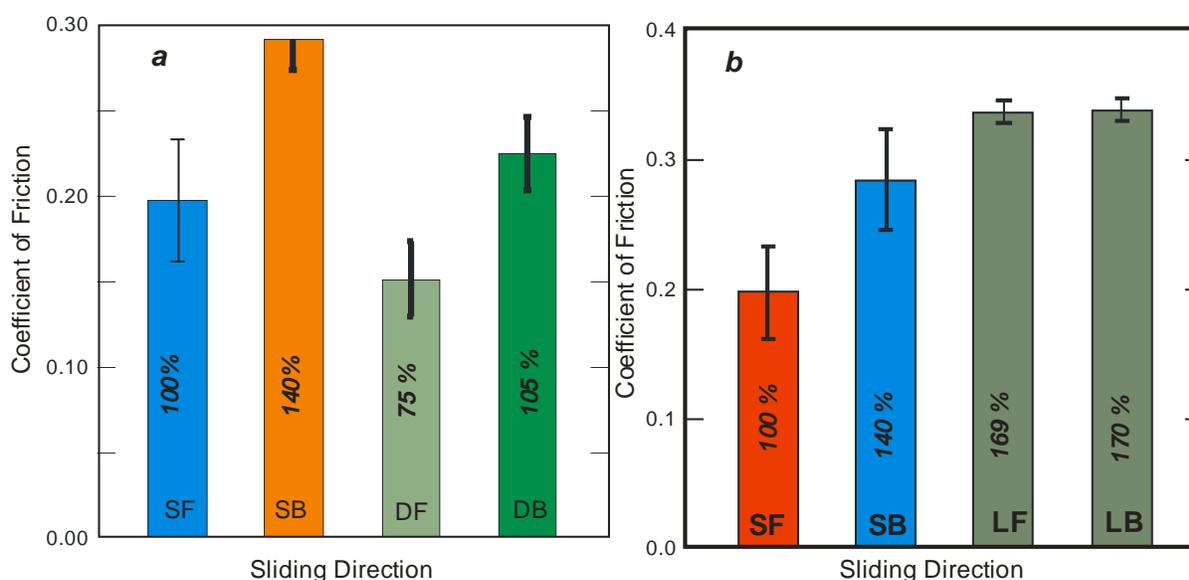

*Figure 7 Coefficient of friction values for each of the sliding directions chosen in this work (refer to figure 7 for equivalent directions). Figure 7-a, depicts a comparison between forward and backward values in rectilinear and diagonal motion. Figure 7-b, depicts a comparison between forward and backward values for rectilinear and lateral motion. In all figures Rectilinear forward motion SF is taken as 100%.*

The origin of the frictional behaviour is rooted in the structure of the surface of the snake. In particular, the presence of fibrils within the scales aids the reptile in conditioning its frictional response. The fibrils are asymmetric as shown in figure (4). The slope of fibril tips is gradual in the direction of forward motion and steep in the direction of backward motion. This asymmetrical tip-shape offers directional resistance to backward motion through acting as a ratchet. Thus, the steep slope of the individual fibril tips offers less resistance to the motion of the reptile in the forward direction than that offered in the reverse direction.

The overall structure of the ventral scales affects friction. The connector tissue, which links the scales, acts as a spring that flexes when the animal attempts to move backwards. Flexing dissipates additional locomotion energy generated by the reptile. This renders backward motion infeasible as it becomes cost prohibitive in terms of energy expenditure. Based on the slope argument, which is in line with the explanation offered by Hazel [15], one would expect that the COF in the lateral direction should be of the same order of magnitude in both directions. That is, the COF moving the skin, for example during lateral undulations, from the left lateral side to the right lateral side would be equal to that obtained upon moving the skin in the converse direction. Such a notion is confirmed in the plots shown in figure (6-b).

It is likely that the trends exhibited by the COF (anisotropy) are also a function of the chemical composition of the Öberchautchen layer. This layer is mainly composed of $\beta$-keratin, which is lamellar. Other natural materials such as camel and horsehair, which have the same keratin, composition are known to exhibit a so-called Differential Friction Effect (DFE) [18]. In DFE the frictional work required by a fibre to slide over another fibre is greater in the direction of tip-to-root than the converse. However in the absence of additional data a conclusion cannot be reached.

### 5.3.2 Acoustical emission

Figure (8) presents a summary plot of the roughness-induced acoustical emission in the directions of sliding chosen in this work. Figure 8-a, shows the repeatability of the raw data whereas figure 8-b, depicts the details of one extracted acoustical fingerprint of the surface. Figure 8-c, meanwhile, presents a comparison plot of the extracted average sound pressure level, in decibels (dB), for the respective sliding directions.

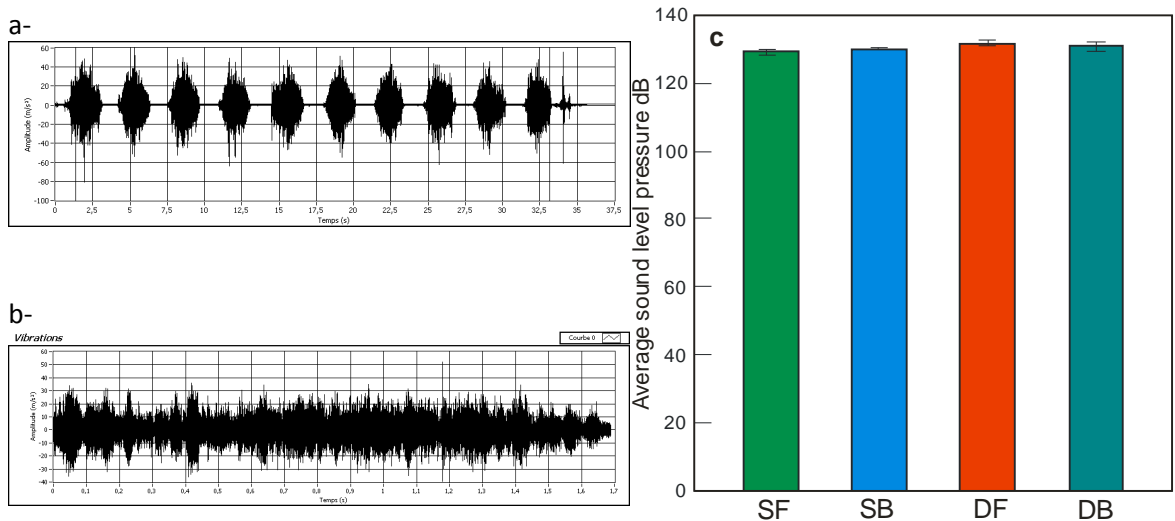

*Figure 8 roughness induced acoustical emission due to friction of the reptile. a- repeatability of experimental runs, b- detailed acoustical vibration fingerprint, and c- average pressure of the radiated sound resulting from the rectilinear and the diagonal motion*

It is noted in figure 8-c that the deduced pressure levels are almost equal in all directions. This is counterintuitive and is in contrast to the behavior of the COF. In particular, one would expect that, in line with the frictional behavior, the sound pressure level in forward motion will significantly differ from that in backward motion. Accordingly, the frictional power dissipated as acoustical power will also depend on the direction of sliding.

Friction of objects is a dissipative process. The energy transferred to the rubbing material is transformed into several forms. One of such forms is sound generation. The intensity of the this generation is a function of the state of the surface as well as of the contact conditions (speed, load, temperature etc.,). Three elements essentially contribute to the production of friction-induced sound. These are relative motion between the contacting solids, vibration of one of the moving solids, and the interface through which the friction energy is dissipated.

In friction practice, energy conversion and dissipation are synergetic events. Each of these processes manifests itself at a different scale. The magnitude of either process (dissipation or conversion) depends on the evolution of instabilities during friction.

Existence of gradients of the thermodynamic forces (pressures, temperatures, stresses, chemical potentials, etc.) is a characteristic of tribo-systems. These gradients drive the transport of energy within the system or between the system and its' surroundings. In this sense, tribo-sytems are considered as energy driven systems. The natural tendency of this class of systems is to operate within a quasi-equilibrium state in the thermodynamic sense. Such a state is facilitated through the construction of an optimal- energy-flow-network that allows the system to redistribute its inherent imperfections (i.e., the internal resistance to the flow of energy). This allows minimization of the different gradients existing between the various components and between these components and the surroundings. In this sense stability of function is a byproduct of the balance between the input energy supply, and the ability of the system to dissipate the supplied energy. When the input energy exceeds the dissipation ability of the system instabilities take place. Instabilities, in turn, give rise to friction excited vibrations (among other occurrences) which are the main sources of sound emission associated with sliding.

Two categories of friction-induced noise are generally identified [19] depending on the nature of instabilities encountered in rubbing. In heavily loaded contacts a type of mechanical instability (so-called stick-slip instability) arises. This is conducive to intense acoustical emission in the form of squeals (such as that encountered in brakes). Under light contact conditions, however, a different form of emission takes place. This is mainly caused by the impact of asperities and roughness features of surfaces. This noise emitted in such a case is characterized by a wideband noise (almost white noise). This is the type of emission encountered in the sliding of the skin which is the subject of this work.

A generalized relationship that relates the acoustic power to the condition of the surface and the sliding speed may be written as [20]:

$$Pa \propto Ra^{\alpha} V^{\beta} \qquad (2)$$

where $\alpha$ and $\beta$ are constants near unity, Ra, is the mean arithmetic value of roughness and V, is the sliding velocity. Equation (2) implies the balanced dependence of the acoustic power on the sliding velocity and the roughness. The roughness parameter Ra may be replaced by Rz, or Rq without altering the form of the dependency [20]. As such interpretation of the results of figure (8-c) has to be attempted in light of the R family of surface roughness parameters. These are summarized in table (1) where the parameters where evaluated from WLI scans in the respective sliding directions.

**Table 1: Summary of surface texture parameters of the skin in the directions of friction measurements.**

| *Direction* | DB | DF | Lateral | Rectilinear |
|---|---|---|---|---|
| **Rz** ($\mu$m) | 0.9 | 0.695 | 0.729 | 0.787 |
| **Ra** ($\mu$m) | 0.147 | 0.138 | 0.111 | 0.115 |
| **Rq** ($\mu$m) | 0.192 | 0.162 | 0.145 | 0.146 |

Equation (2) implies that both the sliding speed and the roughness texture parameter are the main influences in acoustical emission. Assuming, that the constants $\alpha$ and $\beta$ are unity, the maximum power acoustical power dissipated will be given by:

$$Pa = \kappa V Ra = \eta Ra \qquad (3)$$

Where $\eta$ is a constant that accommodates both: the velocity, $V$, (when the velocity of sliding is the same for all scanning directions) and the proportionality constant $\kappa$. So, if the scanning of the surface is performed at the same speed for all directions, then the only parameter that will contribute to the radiation of acoustical power in sliding will be the surface roughness parameter (depending on the choice of a particular R parameter). Referring to table (1) it will be noticed that all profile parameters, for the particular region of measurements, are practically invariant. They don't display appreciable difference in the respective directions. This implies that if the velocity remains the same, the radiated acoustic power will also remain practically directionally invariant. That is, for the same speed of sliding the radiated acoustic power will not depend on the direction of sliding.

The relationship between the radiated acoustic power and the sound pressure level (the measured quantity in figure 8-c) is logarithmic [21]. One of the suggested forms that relate change in sound pressure level (SPL), $Lp$, to the state of roughness is given by,

$$\Delta SPL = Lp - P_{ref} = 20 \left[ \log_{10} (R_x) - \log_{10} (P_{ref}) \right], \quad dB \qquad (4)$$

Where: $P_{ref}$ is the so called reference sound pressure RSP ($2*10^{-5}$ Pa). This quantity is equivalent to the lowest sound pressure that a human ear can hear. The subscript $x$ in equation (4) may be replaced by a, z, or q depending on the texture parameter of choice. Equation (4) may be used to compare the measurements presented in figure (8-c). Thus, by substituting the roughness values from table (1) into equation (4) a map that compares between the measured sound pressure level and the predicted theoretical values may be obtained. Figure (9) presents such a map. Note that the scale in this map is rather deceptive. The values of the SPL are plotted with taking into account the value of the RSP. Thus the map shows the dB value equivalent to the absolute pressure value. Again, it is noticed that the theoretical and the measured values of the SPL are in good agreement. Moreover, the use of the any of member the *R* roughness family yields a reasonable estimation of the SPL. That is, the R values are may be used interchangeably to predict the SPL.

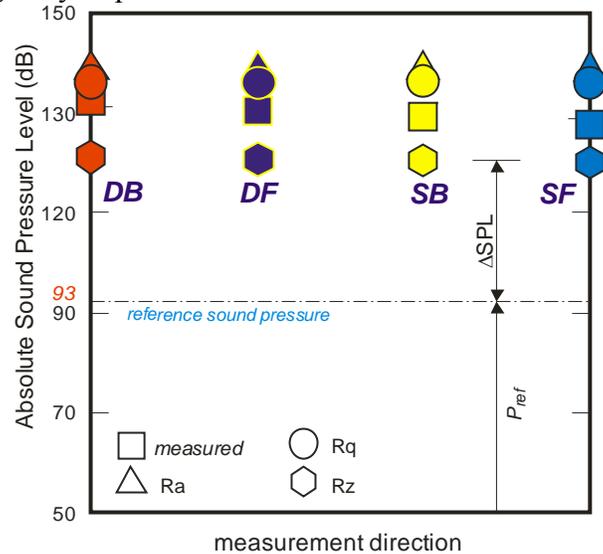

*Figure 9 Comparisons between the theoretical and the measured sound pressure levels in the different sliding directions. Calculated values were obtained by using the different R roughness parameters listed in table 1. Values are plotted taking into account the value of the reference sound pressure (≈ 93 dB).*

## 5.4 Limitations of the study and future work

This study reported the results of an exploratory investigation of the friction response of reptilian shed skin in dry sliding. To the knowledge of the authors, this is the first time in tribology literature that a correlation between surface metrology and frictional behavior for reptiles is introduced. The importance of the study drives from the potential influence reptilian locomotion can exert on manmade surfaces. In particular how can reptilian skin inspire the implementation of smart self tuning tribo-surfaces. Natural surfaces have been a subject of study in recent years. Many have studied plant surfaces (especially lotus leaves) with the aim of developing super-hydrophobic or self cleaning surfaces. Reptilian skin holds the advantage of being a naturally dynamic surface. The skin, in contrast to static plant surfaces, is intended for motion in different environments that are mostly tribologically hostile (abrasive, extremely rough, ultra-dry, etc.,). The initial findings of the study have raised several questions concerning the physics and mechanics of tribological function of the skin.

It was shown that frictional anisotropy is a characteristic of the skin. Preliminary results linked the shape of the fibril tips to anisotropy hence a geometrical origin of frictional anisotropy was implied. Nevertheless, data that would eliminate a chemical origin, based on elemental structure, of the anisotropy is still needed. Thus, it is desirable to obtain friction data under a matrix of nominal loads and sliding speeds. The skin, being essentially a visco-elastic material should exhibit frictional dependence on external load and speed. If the range of variation of the COF is minimal fibril-tip asymmetry would be more of an explanation. Yet, if the adhesive component of the COF shows strong dependence on sliding conditions then the keratinous nature of the Öberchautchen layer would be favored as an explanation. Generation of such data is currently under planning.

The geometrical metrology of the ventral scales indicated that both the macro and micro features of the ventral side are non-uniform. Fibril density, FAR, and spacing between rows were shown to change according to location within the body. Surface areas of individual scales, their aspect ratio and curvature, were also shown to vary along the body. Such features should affect the contact mechanics of the reptile. As a consequence, different parts of the body will sustain different frictional tractions. Moreover, since the area of contact will also differ, the COF may differ not only according to direction but also according to the region of measurements. That is a difference in the COF between skin located at the front and the trailing parts of the trunk. The existence of such a difference was not investigated in this work. The relation between frictional responses and the damage patterns (or lack thereof) in reptilian skin is another point worthy of investigation. One of the explanations for wear resistance of snake skin is the existence of a gradient in the mechanical properties of the different skin layers [22]. From a contact mechanics point of view, this translates into the change of the stiffness of the contact spot. The variation in contact stiffness is augmented by the nature of locomotion in snakes. Snakes generate motion by contraction and relaxation of appropriate muscle groups. This, in theory at least, should cause a variation in the frictional response and consequently affect damage response.

Finally, friction-induced acoustic dissipation is a curious phenomenon in snake locomotion. The results reported in this work show, counter to intuition, that the acoustical pressure is practically invariant in all directions. Acoustics of sliding are normally a function of the roughness state of the surface. Previous measurements [7, 13] indicated that surface texture parameters (the $R_x$ family), which are directly related to change in SPL, vary by region on the body. Acoustical dissipation should also vary according to location within the trunk. More interesting, however, is the behavior of the acoustical power radiated during sliding. In particular, it is desirable to establish what percentage of frictional power is likely to dissipate as sound and which direction is favored in light of the frictional anisotropy exhibited by the skin.

**Conclusions**

This work reported on the frictional behavior of reptilian shed skin obtained from a ball python (python regius). The study intended to link the micro-structure of the skin to the frictional behavior and locomotion of the reptile. Although preliminary, in scope, the results pointed out several interesting features concerning the tribological response of the skin.

The friction profiles of the shed skin are anisotropic. The coefficient of friction exhibits dependence on the direction of sliding. In all measurement directions used in this work backwards motion, defined as going from head-to-tail, resulted in a higher coefficient of friction that that obtained in the converse direction. The trend was also the same for diagonal motion. Directional dependence of friction was not observed, however, in lateral motion. Measurements taken in moving from the right to the left hand side of the ventral scales didn't vary, appreciably, from those taken in motion in the opposite direction.

Study of the profile of the fibril tips suggest that the observed frictional anisotropy is of geometrical origin. This lends support to the view that the fibril tips acts as ratchets that prevent backward motion for the reptile (as it becomes cost prohibitive energy wise). Albeit, no conclusive evidence were forwarded to completely discard the possibility of a chemical origin of the frictional anisotropy. Further studies are needed to clarify this point.

Acoustical emission, associated with friction was invariant to the direction of sliding. This contrasts the behavior of the frictional coefficient. The explanation proposed for this behavior was formulated on the basis of the skin metrology. Acoustic emission due to friction is related to the $R_x$ family texture parameters. For the regions, where the measurements took place, the texture parameters were practically invariant. Sound emission, thereby, was also invariant. Further investigation on the relation of acoustical emission to overall frictional power dissipation was suggested as well.